# Multimodal Recommender System in the Prediction of Disease Comorbidity

Aashish Cheruvu
Central Bucks High School South
aashcheruvu@gmail.com

*Abstract—* **This paper presents two novel applications of deep learning algorithm-based recommender systems, Neural Collaborative Filtering (NCF) and Deep Hybrid Filtering (DHF), for disease diagnosis. Two datasets, first dataset with all diseases and the second dataset with 50 most commonly occurring diseases, were derived from the MIMIC database. The testing, validation and training accuracy of the model with reduced dataset (50 ICD code) was lower (~ 80%) than the model trained on all ICD-9 codes (~ 90%). The model using all ICD codes performed better (80%), also in terms of hit ratio@10, compared to the model with 50 ICD-9 codes (35%). Reasons for a superior performance with dataset using all ICD can be mainly attributed to the higher volume of data and the powerful nature of deep learning algorithms. Compared to literature reports, the novel approach of using deep recommender systems performed well. Results from the deep hybrid filtering model show better in training accuracy (93.75%) compared to NCF model (90.82%), indicating that the addition of text data from clinical notes provided improved performance in predicting comorbidity. The deep learning-based recommender systems have shown promise in accurately predicting subject disease co-occurrence.**

*Keywords—Deep Learning, Recommender Systems, Comorbidity, Natural Language Processing*

## I. Introduction

With the fastest growing demographic of people older than 65, there is a growing incidence of disease comorbidity. To address the comorbidity and the ensuing health care costs, there is a need to develop tools that can learn from the multitude of data available from various sources (viz. Electronic Health Records (EHR), Health insurance claims, etc). Among these various resources, EHRs are one of the most valuable resources, as they provide data from real world evidence. Furthermore, the data in EHR includes information from varied sources including both structured data (e.g., blood pressure, dementia score etc) - which are directly interpretable, and unstructured data (images- CT scan, MRI and text- clinical notes etc.) – that need expert interpretation. With the availability of these vast amounts of data, there is a need for developing tools that can effectively handle the data. The current work used MIMIC-III [1] a publicly available EHR database containing information about 40,000 ICU patients. The databased is organized to include various diseases (included as International Classification of Diseases – ICD-9 codes [2]), clinical notes, CT scans, physiologic signals data etc. To this end Machine Learning (ML) methods that can process large amounts of data in multiple dimensions are promising tools in assisting physicians. ML algorithms can discover, classify, and identify patterns and relationships between various disease characteristics and effectively predict future outcomes of disease. Among the various ML approaches Deep Learning (DL) approaches have shown promise in working with data from multiple modalities including text, image, voice, video etc. Deep learning has had tremendous success across multiple domains including image recognition or natural language processing.

Recommender systems (RS) are algorithms aimed at suggesting relevant items to users (movies to watch, text to read, products to buy etc). RS have the potential to handle the increasing "information overload" effectively and tailor content precisely. RS typically fall into 3 broad categories – content-based filtering, collaborative filtering, and hybrid filtering (a combination of the first two). This paper introduces two RS (collaborative and hybrid filtering) that use state-of-the-art techniques, which were evaluated in predicting the probability of subject disease co-occurrence.

The contributions of this study are three-fold. The first contribution is the development of a hybrid recommender system that combines deep learning based collaborative filtering techniques and text features derived from clinical notes using natural language processing (NLP) techniques in predicting disease co-occurrence. The second contribution is to understand the impact of using partial data (top 50-ICD codes) versus using the sparser full dataset on the accuracy of predicting disease co-occurrence. The third contribution is to understand the impact of changing the proportion of clinical notes on the accuracy of disease co-occurrence prediction.

## II. Related work

While much work has been done on using data mining and in particular NLP/text mining algorithms for the purpose of ICD-9 code prediction, not much work has been done in using recommender systems for ICD code prediction. However, Davis et al., created a rudimentary collaborative RS using cosine similarity for ICD code prediction [3]. While this work makes use of basic techniques, many new techniques have been introduced in the literature that allow for much more accurate models, most notably deep learning-based RS. Also, only collaborative filtering is used for disease prediction, which makes it unable to incorporate useful content related features. In regard to text-based ICD prediction models, Zhang et al., have trained a BERT (bidirectional encoder representations from transformers) model using text data for the purpose of ICD code classification on a proprietary dataset, achieving high performance [4]. However, the authors have not made use of recommender system based methods in disease prediction. Mullenbach et al, have trained a Convolutional Neural Network using text data for ICD code prediction using the MIMIC-III database [5]. Moons et al., have trained a host of deep learning methods using text data and evaluated them using the top 50 ICD codes and all ICD codes on the MIMIC-III [6]. Unlike these and other earlier works, this current work attempts to apply the state-of-the-art deep learning architectures in recommender systems (hybrid and collaborative deep recommender systems) and use ICD-9 codes as well as notes (using NLP techniques) for ICD code prediction.

## III. Methodology

As shown in the Figure 1, the first step was to perform SQL queries to extract relevant information from the MIMIC-III database. This was done with the use of the google-cloud-bigquery python package, used in interaction with the MIMIC III Database stored in Google BigQuery. Exploratory data analysis was performed to identify the data characteristics. As a next step the data was preprocessed to extract the relevant data and perform data wrangling. A major part of this step is Negative Example Generation (NEG). After this the models were trained and evaluated on accuracy, Macro F1, AUC, and Hit Ratio. The scikit-learn package was used to calculate metrics (apart from hit ratio). Sections 3.1.1 and 3.1.2 describe each step in more detail.

Figure 1: Methodology overview

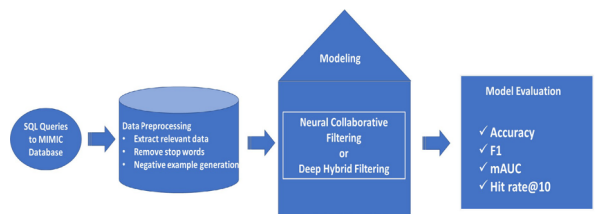

## A. Data Preprocessing

The MIMIC-III dataset is a large dataset relating to patients admitted to critical care units at a large tertiary care hospital. It contains de-identified medical records of patients who stayed from 2001 to 2012 within the intensive care units at Beth Israel Deaconess Medical Center. The database has extensive information on approximately 60,000 ICU admissions at the hospital from 2001 – 2012. Clinical data across many modalities (e.g., image, structured data, clinical notes etc.) are included, but for the task of future disease prediction, the 'DIAGNOSES_ICD' table was used for collaborative filtering and the combination of the 'DIAGNOSES_ICD' table and features derived from the 'NOTEEVENTS' table were used. DIAGNOSES_ICD contains ICD-9 codes (International Classification of Diseases), where individual ICD-9 codes signify specific medical diagnoses. While ICD-9 codes have been used primarily for the purpose of medical billing, standardized "coding" for diseases that is commonly adopted serves very well for the purpose of disease prediction. NOTEEVENTS contains various clinical notes, such as physician notes, radiology notes, discharge summaries, etc. Because discharge summaries report finalized diagnoses from which ICD-9 codes for patients are created, these types of notes were excluded from analysis.

The goal of this study is to explore useful semantic information using both structured and unstructured data. The datasets used were 'DIAGNOSES_ICD' and 'NOTEEVENTS' from which only the free-text clinical note section was used. Furthermore, discharge summary notes were not included as they contain actual ground truth and free-text upon which ICD-9 codes are prepared. Other categories of notes that represent information obtained during the visit were instead utilized. The data were preprocessed to produce separate datasets using two approaches. The first approach is to treat the ICD-9 code independently from each other, find the admissions (unique HADM_ID) for each ICD-9 classification, and consider only records related to the top 50 common ICD-9 codes. The top 50 were chosen because they covered a majority of the dataset: 93.6% of all data. To test the robustness of the model, the next step was to utilize all ICD-9 codes. Evaluations were performed on the two datasets, which will hereby be referred to as top-50-code and all-ICD-codes, respectively.

### 1) Neural Collaborative Filtering Recommender System

The 'DIAGNOSES_ICD' table was used. Label encoding on each subject and each ICD 9 code were performed. Two datasets were created: one with the 50 most commonly occurring ICD codes within the dataset, and another with the full set of ICD codes. A model with all ICD codes (which are relatively sparse data) was used to evaluate the robustness of the model.

An "implicit feedback" RS approach was used. A rating column was added that identified the occurrence of a given ICD-9 code for each subject having a given ICD code as part of the "positive class" (1). However, negative samples are also needed for training the models inorder to maximize the prediction of similar diseases and minimizing in unrelated instances. Subjects who did not have certain ICD codes constitute the "negative class" (0). these co-occurrences had to be generated. Generation was done by randomly sampling subject-disease "pairs", in which a randomly sampled disease was compared with a certain patient's past history. If the disease was not included in the past history, a pair of the subject and disease was created and given a label of 0.

Negative examples found in the dataset were randomly generated at positive class to negative class ratios of 1:10, 1:4, and 1:2. The use of random generation was able to allow for the chance that even when a certain disease is not reported in a subject, it is possible that the subject may have the disease. The model should be able to predict these cases in an accurate way, so random generation, as opposed to incorporating all negative examples, was performed.

### 2) Deep Hybrid Filtering Recommender System

To test the hypothesis for the improvement recommender system performance with the inclusion of symptom information, a novel deep hybrid filtering recommender system was developed. "NOTEEVENTS" transcripts were used for extracting patient symptoms for a given subject. The 'en_core_sci_md' and 'en_ner_bc5cdr_md' models from the SciSpaCy python package were used for implementing the NLP component. SciSpaCy contains SpaCy models trained on biomedical, scientific, and clinical text. The 'en_core_sci_md' model was used to remove i) stop words, punctuation, and other unnecessary information and ii) words that do not have biomedical/clinical significance. Then, the data was passed through the "en_ner_bc5cdr_md" model (a named entity recognition model) was used to isolate tokens (words) that represent symptoms or medications. Because the performance of the NCF model using the top 50 ICD codes was not as good as using all ICD codes, a separate dataset containing only the top 50 ICD codes was not created or used to train the hybrid filtering models. A combination of a subject, an ICD-9 code they had, and a randomly selected symptom mentioned in the text notes constituted individual data points in the "positive class".

NEG for the deep hybrid filtering RS was done by combining a subject, an ICD-9 code they did not have, and a randomly sampled symptoms that the patient had. This was in hopes that this would aid the model in predicting whether a patient had an ICD code using information from whether that symptom was relevant to the ICD-9 code

## B. Model Training and Testing

Two recommender systems that use deep learning approaches were developed sequentially. Firstly, a neural collaborative filtering (NCF) model was used in predicting subject disease co-occurrence, where model derived features for subjects and ICD codes were used. As a next step, a deep hybrid filtering model was created by combining the NCF model developed in earlier step with NLP-derived features (symptoms & medications) from clinical notes (the NOTEEVENTS table). The method of hybridization used to incorporate NLP features was feature combination (Vall et al.). It was hypothesized that the addition of NLP based features improves the model performance in predicting the disease cooccurrence.

*1) Neural Collaborative Filtering Recommender System*

The NCF RS was proposed by Barrett et al [7]. Mathematically, the NCF architecture can be represented as such:

$$\hat{y}_{sc} = f(\mathbf{P}^T\mathbf{v}_s^S, \mathbf{Q}^T\mathbf{v}_c^C \mid \mathbf{P}, \mathbf{Q}, \Theta_f)$$

Here, $\hat{y}_{sc}$ represents a binary prediction of whether there is a co-occurrence of a subject s and an ICD code c. P and Q represent the matrices of subject and ICD code, and v represents the latent/feature vector that $\hat{y}_{sc}$ the model creates. In the models created, an 8-dimensional "embedding" for both P and Q were learned by the NCF models. and theta represents model parameters. Θ represents a binary prediction for the hybrid model. The function f represents function formed by the "dense" neural network layers present in the model, and can be shown as:

$$f(\mathbf{P}^T\mathbf{v}_s^S, \mathbf{Q}^T\mathbf{v}_c^C) = \phi_{out}(\phi_2(\phi_1(\mathbf{P}^T\mathbf{v}_s^S, \mathbf{Q}^T\mathbf{v}_c^C)))$$

where φ represents the function represented by each layer in the neural network, with the subscript denoting the specific layer it represents.

The model architecture for NCF is represented in Figure 2. The inputs to the model are the one-hot encoded ICD-9 code and subject vector. The ICD-9 code vector and subject vector are fed to the user embedding and item embedding respectively, which results in a smaller, denser ICD-9 code and subject vectors. The embedded ICD-9 code vector and subject vectors are concatenated before passing through a series of fully connected layers, which maps the concatenated embeddings into a prediction vector as output. At the output layer, a sigmoid activation function was applied to obtain the most probable class. In the example above, the most probable class is 1 (positive class), since 0.8 > 0.2.

Figure 2: NCF Model Architecture

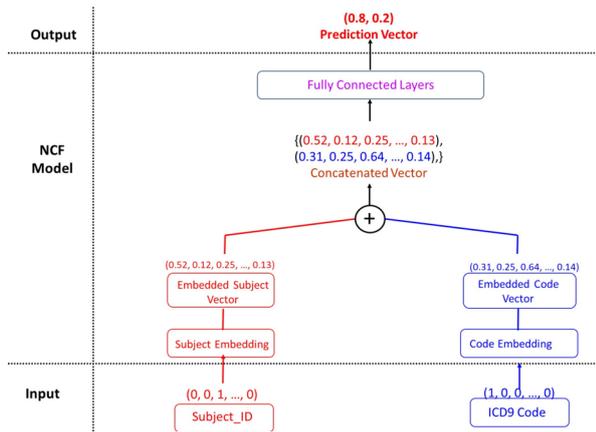

*2) Deep Hybrid Filtering Recommender System*

Just as with the NCF model, embeddings were created for both the subject and ICD-9 code vectors. However, an embedding of the text feature column (which includes both symptoms and medications) was also created. It was hypothesized that creating embeddings for the text feature column in relation to the other columns would help make better predictions.:

$$\hat{y}_{scm} = f(\mathbf{P}^T\mathbf{v}_s^S, \mathbf{Q}^T\mathbf{v}_c^C, R^T v_m^M \mid \mathbf{P}, \mathbf{Q}, \mathbf{R}, \Theta_f)$$

vectors. P, Q and R represent the matrices of subject, ICD code, and symptom. v represents the latent/feature vector that the model creates. In the models created, an 8-dimensional "embedding" for both P, Q and R were learned by the DHF models. Θ represents model parameters. ŷscm represents a binary prediction for the hybrid model. The function f represents function formed by the "dense" neural network layers present in the model, and can be shown as:

$$f(\mathbf{P}^T\mathbf{v}_s^S, \mathbf{Q}^T\mathbf{v}_c^C, \mathbf{R}^T\mathbf{v}_m^M) = \phi_{out}(\phi_2(\phi_1(\mathbf{P}^T\mathbf{v}_s^S, \mathbf{Q}^T\mathbf{v}_c^C, \mathbf{R}^T\mathbf{v}_m^M)))$$

where φ represents the function represented by each layer in the neural network, with the subscript denoting the particular layer it represents.

The model architecture for DHF is represented in Figure 3.

Figure 3: DHF Model Architecture

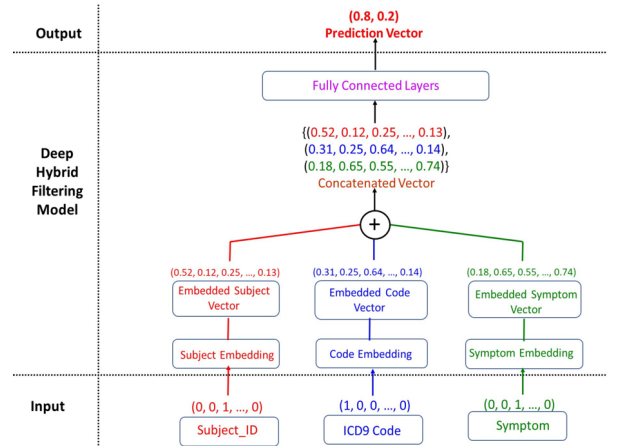

The inputs to the model are the one-hot encoded ICD-9 code vector, subject vector, and symptom vector. The ICD-9 code vector, subject and symptom vector are fed to the ICD-9 code embedding, subject embedding and symptom embedding respectively, which results in a smaller, denser representation of the ICD-9 code, subject and symptom vectors. The embedded ICD-9 code vector, subject and symptom vectors are concatenated before being passed through a series of fully connected layers, which map the concatenated embeddings into a prediction vector as output. At the output layer, the sigmoid function was applied to obtain the most probable class. In the example above, the most probable class is 1 (positive class – subject will develop disease), since 0.8 > 0.2.

## IV. RESULTS AND DISCUSSION

### A. Neural Collaborative Filtering Recommender System

Table 1 presents overall precision performance of models with Top 50 and all ICD codes using NCF. The testing, validation and training accuracy of the model trained on 50 ICD code data was lower ( ~ 80%) than the model trained on all ICD codes (~ 90%). Because of the binary nature of the dataset, test micro F1 was the same as test accuracy. Macro F1 scores for the model with all ICD codes was far higher than with 50 ICD codes. AUC (reflective of model performance) also performed best with model using all ICD codes, with AUC with all ICD codes higher (0.9199) than the AUC with 50 ICD codes (0.6684).

| ICD codes | Accuracy (%) | | | Macro F1 | | | Test AUC | Micro F1 | Hit Ratio @10 (%) |
|---|---|---|---|---|---|---|---|---|---|
| | Training | Validation | Test | Train | Validation | Test | | | |
| Top 50 | 81.45 | 80.36 | 80.37 | 0.6131 | 0.5813 | 0.5207 | 0.6684 | 0.8037 | 35 |
| All | 90.82 | 90.01 | 89.99 | 0.8145 | 0.8323 | 0.8314 | 0.9199 | 0.8999 | 80 |

The model using the 50 ICD codes had a much worse hit ratio of 0.35 as compared to the model with all ICD codes (0.8) for algorithm using all ICD codes reflective performance as a recommender system. The main reason for a superior performance with all ICD compared to 50 ICD codes can be mainly attributed to i) the higher volume of data with dataset containing all ICD codes and ii) the powerful nature of deep learning algorithms. It is intriguing to observe this result, even though there was sparsity making a harder problem to predict.

Table 2 Average Precision Performance with NCF and HCF

| | Accuracy | | | Macro F1 | | | Test AUC | Hit Ratio |
|---|---|---|---|---|---|---|---|---|
| **NCF Models** | | | | | | | | |
| Negative to Positive Ratio | Train | Validation | Test | Train | Validation | Test | | |
| 1 to 10 | 0.9667 | 0.9415 | 0.9411 | 0.4915 | 0.7946 | 0.7936 | 0.9393 | 0.8274 |
| 1 to 4 | 1.0000 | 0.9063 | 0.9048 | 1.0000 | 0.8480 | 0.8453 | 0.9390 | 0.8220 |
| 1 to 2 | 0.8889 | 0.8955 | 0.8946 | 0.8393 | 0.8432 | 0.8420 | 0.9350 | 1.0000 |
| **HCF Models** | | | | | | | | |
| **20k Notes** | | | | | | | | |
| 1 to 10 | 0.9333 | 0.8972 | 0.8974 | 0.7321 | 0.8354 | 0.8357 | 0.9378 | 0.8036 |
| 1 to 4 | 0.8889 | 0.8960 | 0.8966 | 0.8000 | 0.8359 | 0.8370 | 0.9367 | 0.7988 |
| 1 to 2 | 0.8462 | 0.8966 | 0.8973 | 0.7833 | 0.8334 | 0.8347 | 0.9366 | 0.8002 |
| **100k Notes** | | | | | | | | |
| 1 to 10 | 0.9546 | 0.9458 | 0.9459 | 0.8872 | 0.8238 | 0.8242 | 0.9508 | 0.8500 |
| 1 to 4 | 0.9130 | 0.9083 | 0.9083 | 0.4773 | 0.8568 | 0.8566 | 0.9503 | 0.8452 |
| 1 to 2 | 0.9524 | 0.8797 | 0.8800 | 0.9482 | 0.8660 | 0.8662 | 0.9456 | 0.8283 |
| **1 million Notes** | | | | | | | | |
| 1 to 10 | 0.9667 | 0.9440 | 0.9440 | 0.9346 | 0.8066 | 0.8069 | 0.9505 | 0.8536 |
| 1 to 4 | 0.8750 | 0.9109 | 0.9108 | 0.8333 | 0.8582 | 0.8581 | 0.9522 | 0.8539 |
| 1 to 2 | 0.9310 | 0.8863 | 0.8866 | 0.9237 | 0.8735 | 0.8738 | 0.9495 | 0.8388 |

Compared to approaches using only NLP algorithms (such as BERT and CNN) by Zhang et al, in ICD-9 code prediction with MIMIC-III dataset, using NCF performed better [4]. The NCF Macro F1 score of 83.14% was higher than the next highest F1 score of 63% using DR-CAML (convolutional neural network with attention and L1 regularization). Also, the NCF Micro F1 score is higher (89.99%) than the best model reported in the literature (63.3%). This could possibly be due to the limitation of the deep natural language processing algorithms using only text data. The proposed deep recommender systems were able to handle the sparse and imbalanced multimodal data efficiently.

### B. Deep Hybrid Filtering Recommender System

Table 2 presents overall precision performance of models using NCF and DHF. 12 hybrid filtering models in total were evaluated across 8 metrics. Within each set of models trained on different numbers of notes, it was observed that a larger positive class to negative class ratio of 1:10 improved performance. For example, the 1 million note model with a positive class to negative class ratio of 1:2 had a test accuracy of 88.66% and a hit ratio @ 10 of 83.88%, while the 1 million note model with a positive class to negative class ratio of 1:10 had a test accuracy of 94.40% and a hit ratio @ 10 of 85.36%.

Across note categories, it can be seen that the addition of more notes also generally resulted in greater performance. The 1 million note model performed better than the 100,000-note model, and the 100,000-note model performed better than the 20,000-note model. The best overall model in terms of test accuracy and hit ratio @ 10 was the model with the 1:10 positive class to negative class ratio that was trained with text features derived from the million notes. It can also be observed that some models trained on 20,000 notes and 100,000 note, had a worse performance than the original NCF model – this is hypothesized to occur because the reduction of notes results in the removal of some ICD codes from the data.

The addition of data from clinical notes resulted in improved model predictions, as the increase to 1 million notes resulted in models that had a superior performance than the NCF models. It is thought that with models trained on all notes present in the MIMIC-III database would result in an even greater margin of accuracy.

## V. CONCLUSIONS AND FUTURE WORK

This study evaluated two different recommender systems based on deep learning methods. The recommender system algorithm using Neural Collaborative filtering had shown good accuracy (89.46) and hit ratio (85%) in predicting subject and ICD-9 code co-occurrence. The use of hybrid filtering with the text features improved NCF model performance, with the best model (1:10 with 1 million notes) improving accuracy (94.40%) and hit ratio (85.36%). Both the recommender systems were robust, with high AUC of 93.50% for the NCF and an AUC 95.05% for the 1:10 with 1 million notes model. Based on the results, both the deep hybrid filtering model and Neural Collaborative Filtering model have the potential to improve prediction of comorbidity.